# Volume measurement by using super-resolution MRI: application to prostate volumetry

*Estanislao Oubel, PhD[1], Hubert Beaumont, PhD[1], Antoine Iannessi, MD[2]*

[1] MEDIAN Technologies, Valbonne, France
[2] Centre Antoine Lacassagne,  Nice, France

**Purpose** Accuracy and precision of measurements are important for patient follow up in oncology but, unfortunately, partial volume effects introduce an undesired variability between observers. Super resolution techniques (SR) combine multiple acquisitions of an object into a single image richer in details. Herein, the use of SR for reducing variability is investigated in the specific context of prostate measurements. Prostate is typically imaged by T2-weighted MRI in three perpendicular low resolution images, each of them presenting partial volume effects in the direction of the slice selection gradient. SR techniques allow to combine these images into an image presenting the same level of details in all directions. This is expected to increase the accuracy and reproducibility of volume measurements, which in turn improves other derived measurements like PSA density [1][2]. Figure 1 shows an example of reconstruction and the images used as input.

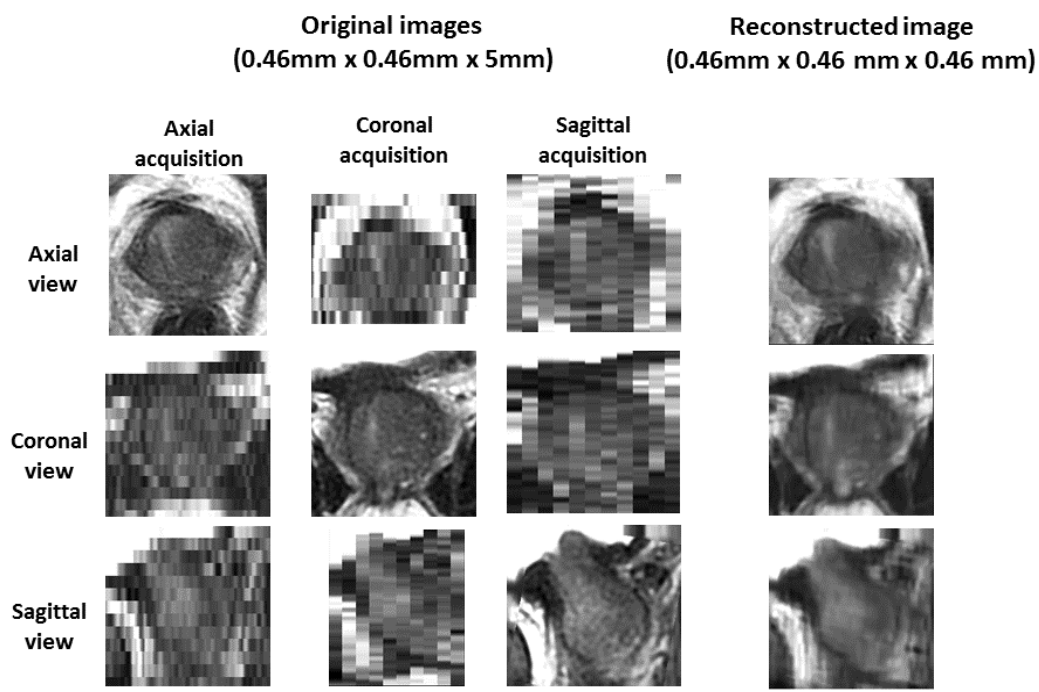

**Figure 1.** Example of images obtained by applying SR techniques to a set of low-resolution perpendicular images.

**Materials and Methods** Experiments were performed on phantom and prostate images. A spherical MRI phantom (General Electric) was imaged to obtain axial, coronal, and sagittal T2 images by using a SE sequence. Image resolution was 0.4mm x 0.4mm x 3mm. Five sets of prostate images freely available from the National Alliance for Medical Image Computing [3] were also employed. Isotropic SR images (1mm x 1mm x 1mm) were created from the three perpendicular acquisitions by using BTK, an open source software originally developed for the processing of fetal MRI images that provides these capabilities [4]. Figure 2 shows a comparison between these two images. Two observers performed repeated volume measurements by using a semiautomatic method (Median Technologies, Valbonne, France) on SR images and original axial acquisitions (N=11 and N=8 respectively). Only one observer segmented prostate images (N=12). A comparison of intra/inter-observer variability for both images was performed by applying a Bartlett's test [5], after verification of normality by using a Shapiro-Wink test [6]. All statistical analyses were performed by using R [7].

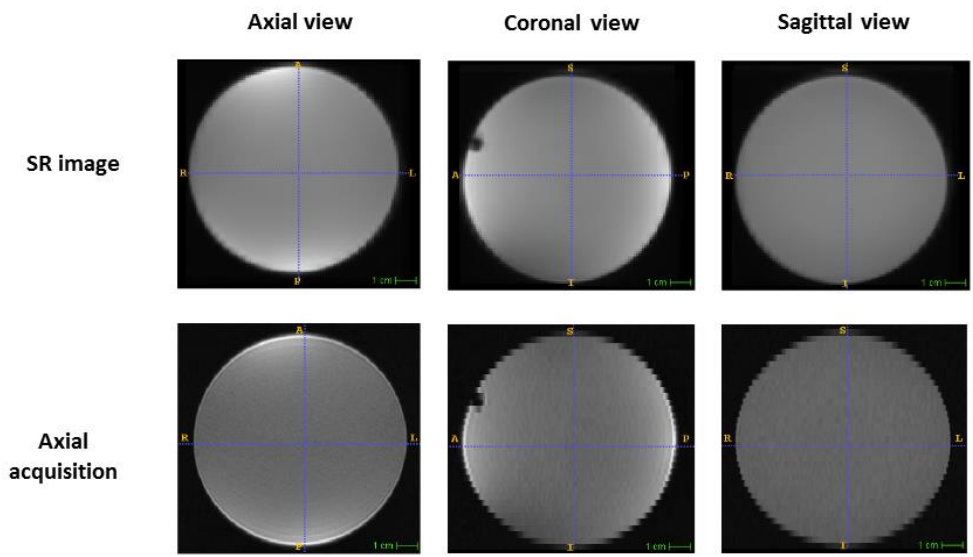

**Figure 2.** Comparison between reconstructed (top row) and original axial acquisition (bottom row) of the phantom. Differences in resolution can be observed in coronal and sagittal views.

**Results** Figure 3 and Figure 4 show some examples of segmentations of the phantom and prostate images respectively. For phantom images, the intra- and inter-observer variability were significantly lower (p<=0.05) for the SR image (8.81ml vs. 16.17ml and 9.39ml vs. 16.26ml respectively). For prostate images, the relative differences were also lower for the SR image (1.6% vs. 3.1%) with p=0.09. A paired t-test showed a significantly higher volume for this image (p < 0.05).

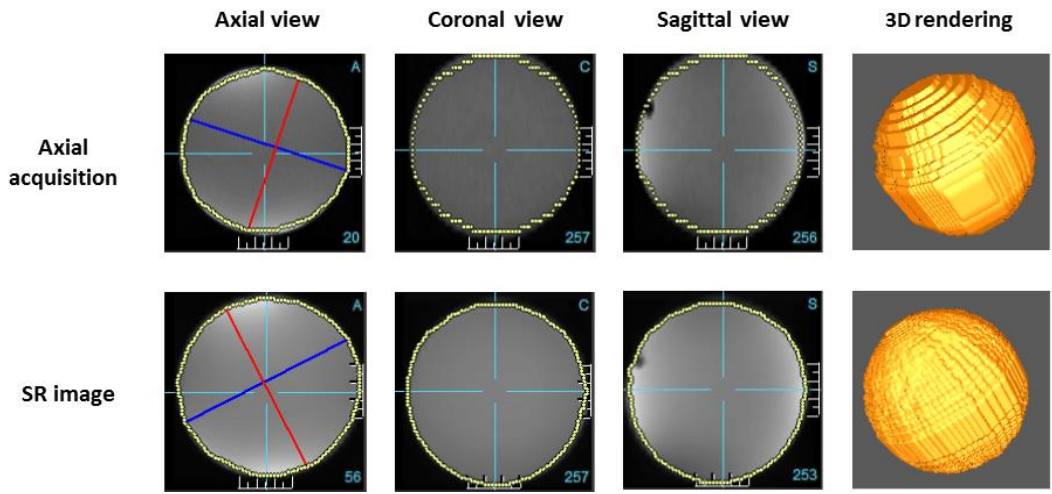

**Figure 3**. Segmentations of the original axial acquisition (top row) and the reconstruction (bottom row) of the phantom. Blue and red lines show respectively the longest and the shortest diameters of the sphere. 3D renderings show that SR images provide better approximations of the real shape.

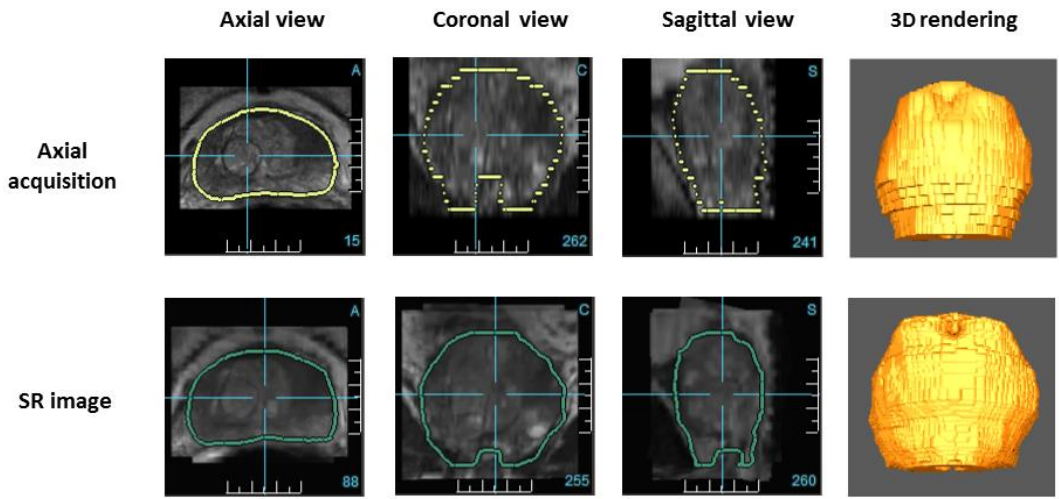

**Figure 4**. Segmentations of the original axial acquisition (top row) and the reconstruction (bottom row) of a prostate.

**Conclusions** SR allowed reducing the variability of volume measurements in phantoms and prostate images. This method could also be applied for lesion measurement, which would be of great importance in clinical trials. Finally, the presented technique allows to obtain images of arbitrary orientations from only three perpendicular acquisitions, which may help medical doctors provide a diagnosis.